\documentstyle[11pt]{article}
\begin{document}\openup6pt

\date{}
 
\title{Maximum mass of a cold compact star}
\author{
{\bf R Sharma\thanks{ranjan$_{-}$sh@hotmail.com}}\\
St. Joseph's College, Darjeeling - 734 104, India\\ \\
{ \bf S. Karmakar\thanks{karma@iucaa.ernet.in} ~and~ \bf S Mukherjee\thanks{sailom@iucaa.ernet.in}} \\
Department of Physics,
North Bengal University,\\
Darjeeling 734 430, India\\
}

\maketitle

\begin{abstract}
We calculate the maximum mass of the class of compact stars described by Vaidya-Tikekar \cite{VT01} model. The model permits a simple method of systematically fixing bounds on the maximum possible mass of cold compact stars with a given value of radius or central density or surface density. The relevant equations of state are also determined. Although simple, the model is capable of describing the general features of the recently observed very compact stars. For the calculation, no prior knowledge of the equation of state (EOS) is required. This is in contrast to the earlier calculations for maximum mass which were done by choosing first the relevant EOSs and using those to solve the TOV equation with appropriate boundary conditions. The bounds obtained by us are comparable and, in some cases, more restrictive than the earlier results.
\end{abstract}

$Keywords:$ Compact star, General Relativity, Strange Star, Maximum mass.
 
$PACS~ Nos.:$ 26.60.+c, 97.60.Jd

\section{Introduction}
In relativistic astrophysics the problem of maximum allowed mass of a very compact star is an important issue. The upper bound on mass, in the case of an ultra-compact object, is crucial for making the distinction between a black hole and a `normal' star. However, the estimation of limiting values of masses is very much model dependent. The maximum mass-radius calculation is essentially based on the assumed equation of state(EOS), although some physical processes which can lead to the instability of the star should also be taken into account. Majority of calculations, based on different neutron star EOS, put a constraint on maximum allowed mass within the range $M \sim (1.46 -2.48)~M_{\odot}$ and radius $b \sim (9-11.7)~ km~$ (see Table 2 of ref. \cite{Haensel01}). However, one can show that above a certain density, it is possible to put a constraint on the maximum allowed mass irrespective of the exotic nature of the EOS used to describe such stars \cite{Shapiro}. For a non-rotating neutron star, Rhoades and Ruffini \cite{Ruffini} have shown that this maximum mass can not be larger than $\sim 3.2~M_{\odot}$. However, data obtained from new generation satellites like `Chandra' predict much lower masses for pulsars than predicted theoretically in the framework of neutron star EOS. Thorsett and Chakrabarty \cite{Thorsett} have pointed out that masses of the majority of the pulsars lie within a narrow range of $(1.35 \pm 0.04)~M_{\odot}$ which also needs an explanation. The new class of compact stars may confirm the conjecture made independently by Witten \cite{Witten} and Farhi and Jaffe \cite{Farhi} that quark matter (and not $^{56}Fe$) is the true ground state of hadrons. Recent investigations show that a number of pulsars like Her X-1 \cite{Li1}, 4U 1820-30 \cite{Bombaci}, \cite{Dey1}, SAX J 1808.4-3658 \cite{Li2}, 4U 1728-34 \cite{Li3}, PSR 0943+10 \cite{Xu}, \cite{Xu1} and RX J185635-3754 \cite{Li2}, \cite{Pons}, earlier thought to be neutron stars, are actually good strange star candidates. Kapoor and Shukre \cite{Kapoor} claim that none of the modern neutron star EOS is good enough to explain the observed mass-radius constraints of known pulsars. 

For more compact (strange) stars, one may make use of the MIT Bag model. The EOS for strange stars made of $u$, $d$ and $s$ quarks, can be written as \cite{Witten}, \cite{Farhi}
\begin{equation}
p = \frac{1}{3}\left(\rho-4B\right), 
\end{equation}
where $\rho$ is the energy-density, $p$ the pressure, $B$ the Bag constant. In the framework of the phenomenological bag model, Banerjee {\em et al} \cite{Banerjee} have obtained a kind of Chandrasekhar limit on maximum mass for strange stars from a consideration of the binding energy. The maximum mass obtained in their calculation shows a scaling with respect to the Bag constant B. Recently, Dey {\em et al} \cite{Dey1} have obtained new sets of EOSs for strange matter based on a model of interquark potential which has the following features: (a) asymptotic freedom, (b) confinement at zero baryon density and deconfinement at high baryon density, (c) chiral symmetry restoration and (d) gives stable uncharged $\beta$-stable strange matter. These EOSs have later been approximated to a linear form by Gondek-Rosi\'{n}ska {\em et al} \cite{Gondek} which has the from,
\begin{equation}
p=a(\rho - \rho_{0}),
\end{equation}
where $\rho_{0}$ denotes the energy density at zero pressure and $a$ is a constant. Making use of equation (2) Harko and Cheng \cite{Harko01} and Zdunik \cite{Zdunik} have shown that for $\rho_{0} = 4B~(B= 56~MeV~fm^{-3})$, the maximum mass of a strange star is $M_{max} = 1.83~M_{\odot}$. Interestingly the maximum mass and radius of a strange star is very much similar to that of a neutron star, the difference being in their mass-radius curves \cite{Alcock1}, \cite{Alcock2}, \cite{Alcock3}. 

Naturally one may ask if it is possible to put constraints on the maximum allowed masses of ultra-compact stars without going into the details of the complex EOSs governing the configurations of such stars. The purpose of this paper is to investigate this problem for a class of stars based on a model given by Vaidya and Tikekar \cite{VT01}. Exact solutions of Einstein's field equations, compiled by Finch and Skea \cite{Finch} and Delgaty and Lake \cite{Delgaty}, reveal that although we have a large abundance of exact solutions (more than a thousand!), most of the solutions are not suitable for the description of realistic stars. In some earlier works \cite{RS01}, \cite{RS02}, it has been shown that the general solution obtained by Mukherjee {\em et al} \cite{Mukherjee} for the class of stars described by Vaidya-Tikekar \cite{VT01} model, is very much relevant for the description of superdense stars. Vaidya-Tikekar \cite{VT01} model has already been studied by many workers (e.g., \cite{Tikekar}, \cite{Patel},\cite{Tikekar02}, \cite{Maharaj}, \cite{Knutsen}) for describing ultra-compact stars. In particular, Knutsen \cite{Knutsen} for a specific value of a  parameter (e.g., $\lambda=2$ in equation (5)), has shown that the maximum mass of a relativistic fluid sphere can not exceed $3~M_{\odot}$ for Baym {\em et al's} \cite{Baym} EOS. Using the solution obtained by Mukherjee {\em et al} \cite{Mukherjee}, which is valid for all physical values of the parameter $\lambda$, it is possible to calculate analytically the maximum possible masses for these class of stars, given the necessary inputs. The model, though given in terms of simple analytic expressions, has been found to be able to describe the essential features of realistic compact stars. In particular, we have shown  that the model can be used to describe pulsars like Her X-1 and SAX J 1808.4-3658 \cite{RS01}, \cite{RS02}. 

The presentation of the paper is as follows. In section 2, we give an outline of the model which will be relevant for the analysis of our work. In section 3, we obtain bounds on the maximum mass for given values of radius or central density or surface density of the compact star. The results obtained are discussed and compared with earlier results in the last section.

\section{Vaidya-Tikekar model}
We consider a static, spherically symmetric star whose interior metric is given by 
\begin{equation}
ds^2 = -e^{2 \gamma(r)} dt^2 + e^{2\mu(r)} dr^2 + r^2(d\theta^2 + 
\sin^2 \theta  d \phi^2).
\end{equation}
The matter within the star is assumed to be of perfect fluid nature and consequently we choose the energy-momentum tensor in the form   
\begin{equation}
T_{ij} = (\rho + p)u_{i}u_{j} + pg_{ij}, 
\end{equation}
where $u^{i}$ is the 4-velocity of the fluid, $\rho$ is the energy-density and $p$ the pressure.

Starting with the ansatz, given by Vaidya and Tikekar \cite{VT01},
\begin{equation}
e^{2\mu}= \frac{1+\lambda r^2/R^2}{1-r^2/R^2},
\end{equation}

Mukherjee {\em et al} \cite{Mukherjee} made use of the pressure isotropy condition to obtain the other metric function,
\begin{equation}
\psi (z)=e^{\gamma }=A\bigg[{\frac{\cos [(n+1)\zeta +\delta ]}{n+1}}-{\frac{
\cos [(n-1)\zeta +\delta]}{n-1}}\bigg].  
\end{equation} 

In equation (6) $\zeta = \cos^{-1} z $, where, $z^{2}=\frac{\lambda}{\lambda+1}(1-\frac{r^2}{R^2})$ and $n^2=\lambda+2$. 

The energy-density $\rho$ and pressure $p$, in this model, are given by 
\begin{equation}
\rho = {1 \over R^2 (1-z^2)} \bigg[ 1 + {2 \over (\lambda + 1)
(1 - z^2)} \bigg]
\end{equation}                                                             
\begin{equation}
p= - {1 \over R^2 (1-z^2)} \bigg[ 1 + {2z \psi_{z}\over
(\lambda + 1) \psi} \bigg],
\end{equation} 
where $\psi_{z}$ denotes differentiation of $\psi$ with respect to $z$. The total mass of the star for a radius $b$ is given by 
\begin{equation}
M(b)  = {(1 + \lambda) b^3/R^2 \over 2(1 + \lambda b^2/R^2)}.
\end{equation}

The model has four parameters, $A$, $\delta$, $R$ and $\lambda$. The conditions $p=0$ and $\gamma=-\mu$ at the boundary, $r=b$, will determine two of these parameters (say $A$ and $\delta$). We need physical inputs to determine the rest. Assuming that $\lambda$ specifies the EOS for the given star in this model, any one of the following, e.g., the radius ($b$), central density ($\rho_{0}$) or surface density ($\rho_{b}$) will determine the parameter $R$. Thus the model gives a complete description of the star, including the radial dependence of density, eqn.(7) and pressure, eqn.(8), and determines the EOS.

\section{Relativistic bound on mass}
A relativistic star, in general, satisfy the following conditions \cite{Shapiro}:\\
(i) Inside the fluid sphere the velocity of sound 
$0 \leq v_{s}(=\sqrt{\frac{dp}{d\rho}}) \leq 1$. Since we are considering only cold stars, the identification of the velocity of sound with $\sqrt{\frac{dp}{d\rho}}$ (i.e., in principle at zero temperature) is unambiguous.\\
(ii) The fluid sphere should be dynamically stable against small radial perturbations.\\

In addition, for a realistic star, the interior solutions must satisfy certain regularity conditions, namely,\\
(iii) Both energy-density and pressure should be positive in the interior of the star.\\
(iv) Pressure should vanish at some finite distance from the centre of the star, determining the radius of the star.

 We will make use of these principles to find bounds on the stellar configurations in Vaidya-Tikekar model \cite{VT01}. From equations (7) \& (8), we obtain
\begin{equation}
\frac{dp}{d\rho} = \frac{z(1-z^2)^2(\psi_{z}/\psi)^2 - (1-z^2)(\psi_{z}/\psi)}{z(1-z^2)(1+\lambda) + 4z}.
\end{equation}

Condition (i), then, implies
\begin{equation}
\frac{1}{1-z^2}\left[\frac{1}{2z} - D\right] \leq \frac{\psi_{z}}{\psi} \leq \frac{1}{1-z^2}\left[\frac{1}{2z} + D\right],
\end{equation}
where, $$ D = \left[4+\frac{1}{4z^2}+  (1+\lambda)(1-z^2)\right]^{1/2}.$$

Imposition of the condition (iii) demand that $\lambda > -1$ and
\begin{equation}
\frac{\psi_{z}}{\psi} \leq - \frac{(1+\lambda)}{2z}.
\end{equation}

Imposition of the condition (iv) gives 
\begin{equation}
\frac{\psi_{z}(z_{b})}{\psi(z_{b})} = - \frac{(1+\lambda)}{2z_{b}},
\end{equation}
where, $$z_{b}^2 = (\frac{\lambda}{\lambda+1})(1-\frac{b^2}{R^2}).$$ 

In ref.\cite{Mukherjee}, combining the two constraints given by equations (11) \& (12) an effective bound on $\psi_{z}/\psi$ was obtained as
\begin{equation}
\frac{1}{1-z^2}\left[\frac{1}{2z} - D\right] \leq \frac{\psi_{z}}{\psi} \leq - \frac{(1+\lambda)}{2z}.
\end{equation}
Equating the two sides\\ 
(a) at $r=0$ gives $\lambda > 3/17$  and at\\
(b) $r=b$ yields 
\begin{equation}
\frac{b^2}{R^2} \leq 1- {\lambda^2 + 5 \lambda + 12 -
(17 \lambda^2 + 82 \lambda + 129)^{1/2} \over \lambda 
(5 + \lambda)}. 
\end{equation} 

A simple calculation will, however, show that equation (15) is a weak bound and it does not ensure the restriction $dp/d\rho \leq 1$ everywhere. For example, if $\lambda=2$, the above condition implies that $b^2/R^2 \leq 0.5$. However, it can be shown that if $b^2/R^2 > 0.4235$, the causality condition is violated. Therefore, a more stringent condition, necessary and sufficient to satisfy the causality condition should be given. It may be pointed out that out of 127 solutions of Einstein's field equations compiled by Delgaty and Lake \cite{Delgaty} only 16 qualify all the tests discussed in the earlier section. Out of 16, only 9 solutions have sound speed which monotonically decreases with radius, a criterion we intend to adopt. 
\begin{itemize}
\item We assume that the value of $\frac{dp}{d\rho}$ is maximum at the centre. The condition $\frac{dp}{d\rho} \leq 1$ at the centre, then, yields
\begin{equation}
(\frac{\psi_{z}}{\psi})_{z_{0}} \geq \frac{(1+\lambda)}{2\sqrt{\lambda}}\left[\sqrt{\lambda+1} \pm \sqrt{21\lambda+1}\right],
\end{equation}
where, $$z_{0}^2 = \frac{\lambda}{(\lambda+1)}.$$ 

Now for $\lambda >3/17$, positivity of pressure (see equation (8)) demands that $\frac{\psi_{z}}{\psi}$ must be a negative quantity which is possible if we choose only the negative sign in the expression on the right hand side of equation (16), i.e.,
\begin{equation}
\frac{\psi_{z}}{\psi}|_{zo} \geq \frac{(1+\lambda)}{2\sqrt{\lambda}}\left[\sqrt{\lambda+1} - \sqrt{21\lambda+1}\right].
\end{equation}

Again, from equation (6), we have 
\begin{equation}
\frac{\psi{_z}}{\psi} = \frac{(n^2-1)}{\sqrt{(1-z^2)}}\left[\frac{\sin[(n-1)\zeta +\delta]-\sin[(n+1)\zeta +\delta]}{(n+1)\cos[(n-1)\zeta +\delta]-(n-1)\cos[(n+1)\zeta +\delta]}\right].
\end{equation}

Combining equations (17) (with equal sign) and (18) at the centre, one can, thus, determine the limiting value of $\delta$ for a given value of $\lambda$.

\item Corresponding to the limiting value of $\delta$, equation (13) can be used to calculate the maximal value of $b^2/R^2$ for the same value of $\lambda$.

\item From equation (9) the compactness of a star in this model is given by
\begin{equation}
u = \frac{M(b)}{b} = \frac{(1+\lambda)}{2(\lambda + \frac{1}{y})}, 
\end{equation}
where, $y=b^2/R^2$. Clearly, the maximum value of $y$ corresponds to the maximum compactness for a given value of $\lambda$. 
\end{itemize}

The method employed here to calculate the maximum mass can be summarised as follows:

(1) We specify a value of $\lambda$ (note that the same $\lambda$ may give different EOS in different stars).

(2) We assume that the maximum mass corresponds to $(\frac{dp}{d\rho})_{0}=1$, i.e., $\frac{dp}{d\rho}=1$ at $r=0$. This determines $y_{max}=(\frac{b^2}{R^2})_{max}$ and hence the maximum compactness $u_{max}=(\frac{M}{b})_{max}$.

(3) There is still one free parameter $R$. It can be specified by giving the values of one of the three: (i) radius $b$, (ii) central density $\rho_{0}$ or (iii) surface density $\rho_{b}$. The relevant relations, in this model, are:
\begin{eqnarray}
\rho_{0}=\frac{3(\lambda+1)}{R^2},~~~~~~
\rho_{b}= \frac{(1+\lambda)(3+\lambda y)}{R^2(1+\lambda y)^2}.
\end{eqnarray}
The three cases are studied in the following:

\subsection{Maximum mass for a given radius}
Here maximum compactness implies maximum allowed mass within a given radius of a star. For example, if $\lambda=2$ and radius $b=10~km$, the maximum mass is $\approx 2.33~M_{\odot}$ and for values of $\lambda \approx 100$, $M_{max} = 2.45 ~ M_{\odot}$. In Table 1, the values of $(b^2/R^2)_{max}$ , $(M/b)_{max}$ and $M_{max}$ for different  radii are given for different $\lambda$. We have plotted the variation of maximum compactness $(M/b)_{max}$ and $(b^2/R^2)_{max}$ with $\lambda$ in fig.1 and fig.2, respectively. 

\begin{table}
\begin{center}
\begin{tabular}{|l|l|l|l|l|r|}  \hline
$\lambda$   &  $(\frac{b^2}{R^2})_{max}$ & $ (\frac{M}{b})_{max}$ & \multicolumn{3}{|c|}{$M_{max}/M_{\odot}$}\\ \hline
 &  &  &   $b=10~km$ & $b= 8~km$ & $b=6~km$ \\ \hline\hline
1 &  0.4618 & 0.3159 &  2.14 & 1.71 & 1.28 \\  \hline
2 &  0.4234 & 0.3438 & 2.33 & 1.86 & 1.39 \\  \hline
3 &  0.3727 &  0.3519 & 2.38 & 1.90 & 1.43 \\  \hline
4 &  0.3297 &  0.3554  & 2.41 & 1.92 & 1.44  \\  \hline
5 &  0.2944  & 0.3573 & 2.42 & 1.93 & 1.45  \\  \hline
7 &  0.2417  & 0.3591 & 2.43 & 1.94 & 1.46  \\  \hline
10 &  0.1898  & 0.3602 & 2.44 & 1.95 & 1.46 \\  \hline
20 &  0.1102  & 0.3611 &  2.44  & 1.95 & 1.46 \\  \hline
50 &  0.0486 & 0.3614 & 2.45 & 1.96 & 1.47     \\ \hline
100 &  0.0252 & 0.3615 &  2.45 & 1.96 & 1.47 \\ \hline
200 &  0.0128 & 0.3615 & 2.45 & 1.96 & 1.47 \\
\hline
\end{tabular} 
\caption{Maximum mass ($M_{max}$) of a star for different radii $b=10, 8$, and $6~km$ and for  different choices of the parameter $\lambda$.}
\end{center}
\end{table}

\subsection{Maximum mass for a given surface density($\rho_{b}$)}
To find the maximum mass, we write the mass in terms of surface density  
\begin{equation}
M  = {(1 + \lambda)^{3/2} y^{3/2}(3 + \lambda y)^{1/2} \over 2 \sqrt {\rho_{b}}(1 + \lambda y)^2}.
\end{equation}
To obtain the maximum mass of a realistic star one has to put some physically acceptable lower bound on the value of surface density. Rhoades and Ruffini \cite{Ruffini} obtained a maximum neutron star mass $\simeq 3.2 M_{\odot}$ allowing an uncertain nature in the EOS above a fiducial density $4.6 \times 10^{14}gm~cm^{-3}$. If we take $4.6 \times 10^{14}gm~cm^{-3}$ as the surface density in equation (21) we get $M_{max} \approx 3.01 M_{\odot}$ which is comparable with the result obtained by Rhoades and Ruffini \cite{Ruffini}. Kalogera and Baym \cite{Kalogera} obtained an upper bound on the neutron star mass equal to $2.9 M_{\odot}$ regarding the EOS as valid upto twice nuclear matter saturation density, $\rho_{nuc} = 2.7 \times 10^{14} gm~cm^{-3}$. In Table 2 we have shown the maximum possible masses and radii for different choices of surface density.

The EOS obtained by Dey {\em et al} \cite{Dey1} for strange stars, namely SS1 and SS2, give rather low values of maximum mass ($M_{max} = 1.33 M_{\odot}$ for SS1 and $M_{max} = 1.44 M_{\odot}$ for SS2 ) where the approximated linearised EOS has $\frac{dp}{d\rho} = a$, a constant. For SS1, $a = 0.463$ and  $\rho_{b} = 1.15 \times 10^{15}~gm/cm^3$ and for SS2, $a = 0.455$ and $\rho_{b} = 1.33 \times 10^{15}~gm/cm^3$. If we consider the same values of $\rho_{b}$ in equation (21) in our model with $(\frac{dp}{d\rho})_{0}=1$, we get higher values of maximum masses  and radii as $M_{max} = 1.9 M_{\odot}$ and $b = 7.98~km$ for SS1 and $M_{max} = 1.77 M_{\odot}$ and $b = 7.43~km$ for SS2, respectively, where $\lambda=53.34$ for SS1 and $\lambda=230.58$ for SS2 \cite{RS02}. A lower value of $a$ will lower the value of maximum mass possible, vide Table 4. Note that for a large $\lambda$, the EOS in our model is almost linear, giving a constant value for $\frac{dp}{d\rho}$. The variation of $M_{max}$ with $\rho_{b}$ is shown in fig.3. 
  
\begin{table}
\begin{center}
\begin{tabular}{|l|l|l|l|l|l|l|l|l|r|}  \hline
$\lambda$ &  \multicolumn{2}{|c|}{$\rho_{b}=5.4\times 10^{14}$} & \multicolumn{2}{|c|}{$\rho_{b}=10.8\times 10^{14}$} & \multicolumn{2}{|c|}{$\rho_{b}=4.6\times 10^{14} $} & \multicolumn{2}{|c|}{$\rho_{b}=5.1\times 10^{14}$}\\ \hline
&  $M_{max} $ & $b_{max}$ & $M_{max}$ & $b_{max}$ & $M_{max}$ & $b_{max}$ & $M_{max}$ & $b_{max}$ \\ \hline\hline
1 &  2.61 & 12.19 & 1.84 & 8.62 & 2.83 & 13.21 & 2.68 & 12.55 \\  \hline
2 &  2.78 & 11.93 & 1.96 & 8.44 & 3.01 & 12.93 & 2.86 & 12.28    \\  \hline
3 &  2.78 & 11.66 & 1.96 & 8.24 & 3.01 & 12.63 & 2.86 & 12.00   \\  \hline
4 &  2.76 & 11.47 & 1.95 & 8.11 & 2.99 & 12.43 & 2.84 & 11.80 \\  \hline
5 &  2.74 & 11.33 & 1.94 & 8.01 & 2.97 & 12.28 & 2.82 & 11.66 \\  \hline
7 &  2.71 & 11.15 & 1.92 & 7.88 & 2.94 & 12.08 & 2.79 & 11.48 \\  \hline
10 & 2.68 & 11.00 & 1.90 & 7.78 & 2.91 & 11.92 & 2.76 & 11.22 \\  \hline
20 & 2.64 & 10.79 & 1.87 & 7.63 & 2.86 & 11.70 & 2.72 & 11.11  \\  \hline
50 & 2.61 & 10.66 & 1.85 & 7.54 & 2.83 & 11.55 & 2.68 & 10.97    \\ \hline
100 & 2.60 & 10.61 & 1.84 & 7.50 & 2.82 & 11.50 & 2.67 & 10.92\\ \hline
200 & 2.59 & 10.59 & 1.83 & 7.49 & 2.81 & 11.47 & 2.67 & 10.90 \\ \hline\hline
\end{tabular}
\caption{Maximum mass ($M_{max}$) in $M_{\odot}$ and corresponding radius $b_{max}$ in km of a star for different choices of the parameter $\lambda$ and surface density in units of $gm~cm^{-3}$.}
\end{center}
\end{table}

\subsection{Maximum mass for a given central density ($\rho_{0}$)}
The mass in terms of central density can be written as
\begin{equation}
M  = {\sqrt{3} (1 + \lambda)^{3/2} y^{3/2} \over 2 \sqrt {\rho_{0}}(1 + \lambda y)}
\end{equation}
The variation of $M_{max}$ with central density is shown in fig.4. Very little information is available regarding the central densities of realistic compact stars. However, the central density is useful in determining the static hydrostatic stability of realistic stars under radial perturbations. It is known that in the mass-central density diagram, the configurations with $\rho>\rho^{m}_{0}$ ($\rho^{m}_{0}$ corresponds to the maximum mass) are unstable with respect to radial perturbations. Introducing a time-dependence of the perturbations complicates the calculations, but the configuration remains unstable with respect to fundamental modes of pulsation if the EOS is not changed by perturbation. It has been shown elsewhere (\cite{Knutsen}, \cite{RS02}, etc.) that the Vaidya-Tikekar stars are stable against small radial perturbations.

\section{Discussions}

We have calculated in this paper the maximum mass of a class of compact stars satisfying  necessary physical constraints. Results obtained are outlined as follows:

\begin{itemize}
\item We have obtained an upper bound on the compactness ($M/b$) of a star described by Vaidya-Tikekar \cite{VT01} model (see Table 1) for different values of the parameter $\lambda$. The compactness, for large values of $\lambda$ (e.g., $\lambda \sim 100$), becomes almost independent of $\lambda$, the limiting value being $\sim 0.3615$. This is less than the Bondi \cite{Bondi} limit $\sim 0.3906$.  Negi and Durgapal \cite{Negi} have shown that in the case of Tolman's type VII solution, a fluid sphere configuration remains stable for $\frac{M}{b} \leq 0.3428$. Based purely on causality limit, the upper bound on compactness obtained here, lies slightly above this value.  

\item We have obtained a better bound for the radial parameter $b^2/R^2$ as compared to the earlier results obtained by Mukherjee {\em et al} \cite{Mukherjee}. The new bound obtained here is in agreement with the earlier work done by Knutsen \cite{Knutsen} for a specific value $\lambda=2$. 
\end{itemize} 

\begin{table}
\begin{center}
\begin{tabular}{|l|l|l|r|}  \hline\hline
Reference & Model &  $M_{max}/M_{\odot}$  \\ \hline\hline
Ruffini and Rhoades \cite{Ruffini} & Neutron star (causality principle) & 3.2\\ \hline
Haensel {\em et al}\cite{Haensel01} & Neutron star (causality principle) & $3.0(5\times10^{14}~gm~cm^{-3}/\rho_{b})^{1/2}$ \\  \hline
Baldo {\em et al} \cite{Baldo} &Neutron star(BBB2 EOS($npe\mu$)) & 1.92 \\  \hline
Kalogera and Baym \cite{Kalogera} & Neutron star(WFF88 \cite{WFF} EOS) & $2.9$\\ \hline
Witten \cite{Witten} & Strange star( Bag model) & $2.0 (B_{0}/B)^{1/2},~B_{0}=56MeV~fm^{-3}$ \\  \hline
Burgio {\em et al} \cite{Burgio} & Strange star(Bag model) & $1.45 - 1.65$\\ \hline
Banerjee {\em et al} \cite{Banerjee} & Strange star(Bag model) & $1.54~(B^{1/4}=145~MeV$) \\ \hline
Mak and Harko \cite{Mak} & Charged strange star(Bag model) &  2.86  \\  \hline
Harko and Cheng \cite{Harko01} & Strange star(Bag model)  &  1.83 \\  \hline
Cheng and Harko \cite{Cheng} & Strange star(Bag model)  &  2.016 \\  \hline
Knutsen\cite{Knutsen} & Vaidya-Tikekar model \cite{VT01}, $\lambda=2$& $3.0$ \\  \hline
Present work & Vaidya-Tikekar model \cite{VT01}, any $\lambda$, & $3.01$ ($\lambda=2$, $\rho_{b}=4.6\times 10^{14}~gm~cm^{-3}$)\\ 
  & $(\frac{dp}{d\rho})_{0}=1$  & $2.82$ ($\lambda=100$, $\rho_{b}=4.6\times 10^{14}~gm~cm^{-3}$)\\ 
 &  & $2.45$ ($\lambda=100$, $b=10~km$)\\  
 &  & $1.96$ ($\lambda=100$, $b=8~km$)\\  
 &  & $1.47$ ($\lambda=100$, $b=6~km$)\\  \hline 
\end{tabular}
\caption{Maximum mass configurations obtained in different models.}
\end{center}
\end{table}
   
\begin{itemize}
\item It is observed from Table 2 that for a given surface density ($\rho_{b}$), maximum mass first increases and then decreases with $\lambda$. For example, for $\rho_{b}=4.6\times10^{14}~gm~cm^{-3}$, this maxima is obtained when $\lambda \approx 3$ and the corresponding values of maximum mass and radius are $3.01 M_{\odot}$ and $12.63~km$, respectively. 

\item The maximum mass and corresponding radius decreases with the increase of surface density. A comparative study of maximum masses for different models are listed in Table 3.

\item If we consider a class of EOSs so that $(\frac{dp}{d\rho})_{0}= a$,   with $a < 1$, the relevant causality condition, eqn (17) gets modified to
\begin{equation}
\frac{\psi_{z}}{\psi}|_{zo} \geq \frac{(1+\lambda)}{2\sqrt{\lambda}}\left[\sqrt{\lambda+1} - \sqrt{(20~a + 1) \lambda+1}\right].
\end{equation}

 It is seen that the maximum compactness, for a given $\lambda$, decreases as the value of $a$ is lowered. For given values of surface density and $\lambda$, we have calculated the maximum mass and the corresponding radius for different choices of $a$, and these are shown in Table 4. It shows that for a star with an EOS giving $a=0.455$ and having surface density $\rho_{b}=1.33\times 10^{15}~gm~cm^{-3}$, the maximum possible mass will be $M_{max}\approx 1.4~M_{\odot}$ and radius $b \approx 6.6~km$. Although the present model covers only a class of compact stars, the results obtained are in conformity with the general results obtained directly by numerical calculations. Thus, if one chooses $\rho_{b}=4~B$ with the Bag constant $B=10^{14}~gm~/cm^3~(56~MeV~fm^{-3})$ and $a=0.333$ in the Bag model, one gets $M_{max}\approx 2M_{\odot}$ and corresponding radius $b_{max}\approx 11~km$ (vide \cite{Witten}, \cite{HZS86}). If we use the same inputs for $\rho_{b}$ and $a=(\frac{dp}{d\rho})_{0}$ and $\lambda=100$, we get $M_{max}\approx 2.12M_{\odot}$ and corresponding radius $b_{max}\approx 11.78~km$, comparable to the numerical values obtained by \cite{Witten}, \cite{HZS86}.
\end{itemize}

\begin{table}
\begin{center}
\begin{tabular}{|l|l|l|l|l|l|l|l|l|l|r|}  \hline
$a=(\frac{dp}{d\rho})_{0}$ & $ (\frac{M}{b})_{max}$ & \multicolumn{2}{|c|}{$\rho_{b}=4.6\times 10^{14}$} & \multicolumn{2}{|c|}{$\rho_{b}=5.6\times 10^{14}$} & \multicolumn{2}{|c|}{$\rho_{b}=1.15\times 10^{15} $} & \multicolumn{2}{|c|}{$\rho_{b}=1.33\times 10^{15}$}\\ \hline
&  & $M_{max} $ & $b_{max}$ & $M_{max}$ & $b_{max}$ & $M_{max}$ & $b_{max}$ & $M_{max}$ & $b_{max}$ \\ \hline\hline

1     & 0.3615 & 2.82  & 11.50 & 2.55 & 10.42 & 1.78 & 7.27 & 1.65 & 6.76   \\ \hline
0.500 & 0.3206 & 2.47  & 11.38 & 2.24 & 10.31 & 1.56 & 7.19 & 1.45 & 6.69   \\  \hline
0.463 & 0.3129 & 2.40  & 11.34 & 2.18 & 10.27 & 1.52 & 7.17 & 1.41 & 6.67   \\  \hline
0.455 & 0.3110 & 2.38  & 11.33 & 2.16 & 10.26 & 1.51 & 7.16 & 1.40 & 6.66   \\  \hline
0.333 & 0.2658 & 1.98  & 10.99 & 1.79 & 9.96 & 1.25 & 6.95 & 1.16 & 6.46    \\  \hline
\end{tabular}
\caption{Maximum mass ($M_{max}$)in $M_{\odot}$ and corresponding radius $b_{max}$ in km of a star for different choices of the parameter $a$ and surface density $\rho_{b}$ in units of $gm~cm^{-3}$. We assumed $\lambda=100$.}
\end{center}
\end{table}

\begin{itemize}
\item Consideration of rotation and the presence of charge may change this scenario to some extent. However, both the effects will increase the maximum mass limit. For example, Mak and Harko \cite{Mak} have shown that the maximum mass and radius of a charged strange star with a linearized EOS in the MIT bag model are $M_{max}=2.86 M_{\odot}$ and $b=9.46~km$, respectively, where the value of $(dp/d\rho)=1/3$. In the uncharged limit, the corresponding values are $M_{max}=1.9638 M_{\odot}/\sqrt{B_{60}}$ and $b=9.46~km/\sqrt{B_{60}}$, with a scaling with respect to bag constant $B = 60~MeV~fm^{-3}$ \cite{Witten}, \cite{Haensel01}. This shows that in the presence of charge the maximum mass increases which is expected as the gravitational attraction is now partly balanced by the Coulomb repulsion. In an earlier work \cite{RS04}, the solution obtained by Mukherjee {\em et al} \cite{Mukherjee} was extended to the case of a static charged spherical distribution of matter. The solution obtained there may be used to study similar effects of charge on the maximum mass. It can be shown that for a small $\lambda$, both maximum mass and the corresponding radius increase in the presence of charge, e.g., for $\rho_{b}=2\rho_{nuc}$, $\lambda=2$ and $\alpha=0.4$ ($\alpha$ is a measure of charge density \cite{RS04}), we found that $M_{max}=2.87~M_{\odot}$ and $b=11.99~km$, both of which are higher than the values in the uncharged ($\alpha=0$) case, as shown in Table 5. However, if we consider $\lambda=1000$, the  maximum mass and corresponding radius for the same values of $\rho_{b}$ and $\alpha$ are $M_{max}=2.59~M_{\odot}$ and $b=10.57~km$, exactly the same as in the uncharged case. Thus, the effect of charge is very small on maximum mass and it becomes negligible for higher values of $\lambda$ in this model. Similarly, if rotation is taken into account, maximum mass limit is expected to increase marginally as pointed out in ref\cite{Haensel01}. 
\end{itemize}

\begin{table}
\begin{center}
\begin{tabular}{|l|c|c|c|c|c|r|}  \hline\hline
$\lambda$ & $\alpha$ &   $y_{max}$ & $(\frac{M}{b})_{max}$ & $M_{max}(M_{\odot})$ & $b_{max}~(km)$  \\ \hline\hline
2 & 0.0 &  0.4234 & 0.3439 & 2.78 & 11.93\\ 
  & 0.4 &  0.4474 & 0.3529 & 2.87 & 11.99\\ \hline
100 & 0.0 & 0.0252 & 0.3615 & 2.60 & 10.62\\ 
  & 0.4 &  0.0252 & 0.3616 & 2.60 & 10.62\\ \hline
1000 & 0.0 &  0.0026 & 0.3615 & 2.59 & 10.57\\ 
  & 0.4 &  0.0026 & 0.3615 & 2.59 & 10.57\\ \hline  
\end{tabular}
\caption{Maximum mass configurations for a charged sphere of surface density $\rho_{b}=5.4\times 10^{14}~gm~cm^{-3}$.}
\end{center}
\end{table}
To conclude, we have shown that the Vaidya-Tikekar model and its solution of Mukherjee {\em et al} \cite{Mukherjee} provide a simple method of studying systematically the maximum mass problems of compact stars. The calculations can be done without any prior knowledge of the EOS. Although the class of stars described by this simple model may not cover, in principle, all compact stars, it nevertheless captures the gross features of known compact stars, as is evident from a comparison with results obtained by solving the TOV equation with different conventional equations of state for Neutron or strange stars, given in Table 3 and Table 4.

\section*{Acknowledgment}
\textit{We would like to thank the IUCAA Reference Centre, North Bengal University, for providing facilities during the course of this work. A part of the work was done when SM visited IUCAA, Pune, as a visiting Associate.}

\newpage
\begin{figure} 
\begin{center}
\vspace{6cm}
\special{eps: 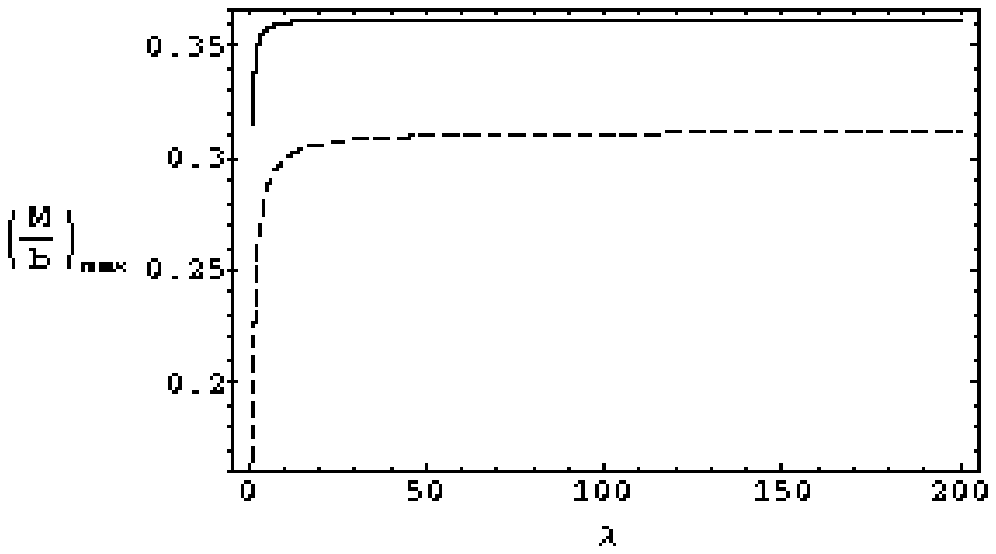 x=10cm y=6cm}
\caption{Variation of $(M/b)_{max}$ with $\lambda$ (solid line for $a=1$ and dashed line for $a=0.455$). }
\end{center} \label{fig.1:}
\end{figure}

\begin{figure} 
\begin{center}
\vspace{6cm}
\special{eps: 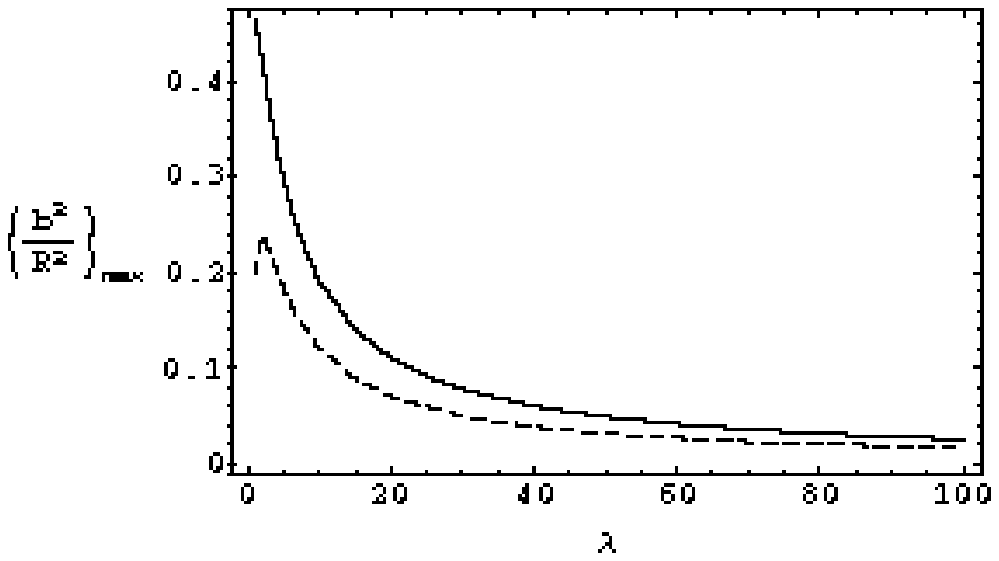 x=10cm y=6cm}
\caption{Variation of $(b^2/R^2)_{max}$ with $\lambda$ (solid line for $a=1$ and dashed line for $a=0.455$).}
\end{center} \label{fig.2: }
\end{figure}
\newpage
\begin{figure} 
\begin{center}
\vspace{6cm}
\special{eps: 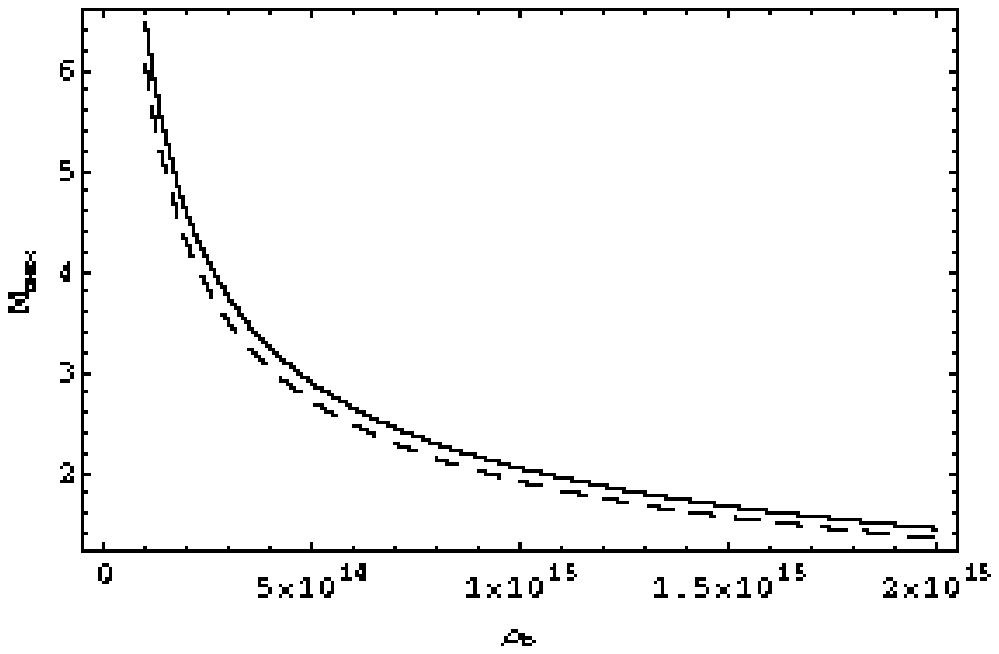 x=10cm y=6cm}
\caption{Variation $M_{max}/M_{\odot}$ with surface density $\rho_{b}$ ($gm/cm^3$) (solid line for $\lambda = 3$ and dashed line for $\lambda = 100$).}
\end{center} \label{fig.3: }
\end{figure}

\begin{figure} 
\begin{center}
\vspace{6cm}
\special{eps: 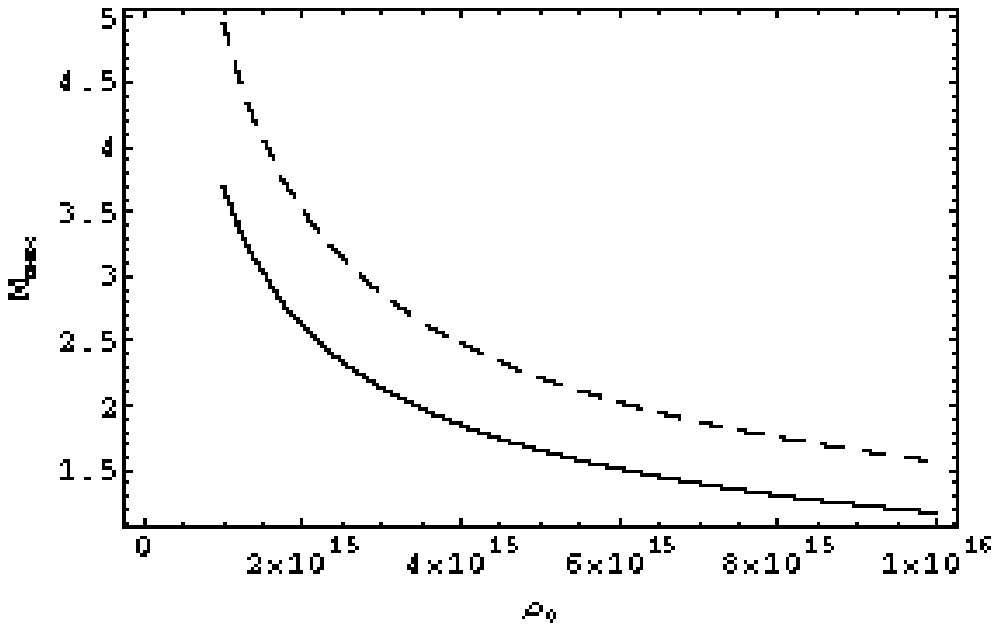 x=10cm y=6cm}
\caption{Variation $M_{max}/M_{\odot}$ with central density $\rho_{0}$($gm/cm^3$) (solid line for $\lambda = 3$ and dashed line for $\lambda = 100$).}
\end{center} \label{fig.4: }
\end{figure}

\end{document}